\RequirePackage{lineno} 
\documentclass[aps,prc,superscriptaddress,nofootinbib,showpacs,floatfix,twocolumn]{revtex4}
\usepackage{graphicx} 
\usepackage{float} 
\usepackage{amssymb}
\usepackage{multirow}
\usepackage{tabularx}
\usepackage[export]{adjustbox}
\usepackage{url}
\usepackage[colorlinks,linkcolor=blue,citecolor=blue,filecolor=black,urlcolor=blue]{hyperref} 
\usepackage{cancel}
\usepackage{MnSymbol}
\usepackage{multirow}
\usepackage{bm}
\usepackage{color}
\definecolor{dgreen}{cmyk}{1.,0.,1.,0.1}        
\definecolor{orange}{cmyk}{0.,0.353,1.,0.}    

\newcommand{\pT} {\ensuremath{p_{\mathrm{T}}}}

\newcommand{\pt}    {p_{\rm T}}

\usepackage{color}
\usepackage{harpoon}
\definecolor{my}{rgb}{1, 0, 0}
\newcommand{\XeXe}{\ensuremath{\mbox{Xe--Xe~}}}
\newcommand{\PbPb}{\ensuremath{\mbox{Pb--Pb~}}}
\newcommand{\mpt}{\mbox{$[p_\mathrm{T}]$}}
\newcommand{\dpt}{\mbox{$\delta p_\mathrm{T}$}}

\newcommand{\la}{\langle}
\newcommand{\lla}{\left\langle}
\newcommand{\ra}{\rangle}
\newcommand{\rra}{\right\rangle}
\begin{document} 

\title{A unified algorithm for multi-particle correlations between azimuthal angle and transverse momentum in ultra-relativistic nuclear collisions}


\author{Emil Gorm Dahlb\ae k Nielsen}
\affiliation{Niels Bohr Institute, Jagtvej 155A, 2200 Copenhagen, Denmark}
\author{Nina Nathanson} 
\affiliation{Niels Bohr Institute, Jagtvej 155A, 2200 Copenhagen, Denmark}
\author{Kristjan Gulbrandsen} 
\affiliation{Niels Bohr Institute, Jagtvej 155A, 2200 Copenhagen, Denmark}
\author{You Zhou}
\email{you.zhou@cern.ch}
\affiliation{Niels Bohr Institute, Jagtvej 155A, 2200 Copenhagen, Denmark}

\date{\today}

\begin{abstract}

Multi-particle correlations between azimuthal angle and mean transverse momentum are a powerful tool for probing size and shape correlations in the initial conditions of heavy-ion collisions. These correlations have also been employed to investigate nuclear structure, including potential nuclear shape phase transitions at the energy frontier. However, their implementation is highly nontrivial, and prior studies have been mostly limited to lower-order correlations, such as the modified Pearson correlation coefficient, $\rho(v_{\rm n}^{2}, [p_{\rm T}])$. This paper presents a unified framework that employs a recursive algorithm, enabling the efficient evaluation of arbitrary-order correlations while maintaining computational efficiency. This framework is demonstrated using widely adopted transport models, including AMPT and HIJING. The proposed unified algorithm for multi-particle correlations between azimuthal angle and transverse momentum provides a systematic and efficient approach for multi-particle correlation analyses. Its application in experiments at the Relativistic Heavy Ion Collider and the Large Hadron Collider facilitates the exploration of nuclear structure at ultra-relativistic energies.

\end{abstract}
\maketitle

\section{Introduction}\label{sec:intro}
Ultrarelativistic collisions of heavy ions at facilities such as RHIC and the LHC create a strongly-coupled, deconfined matter called the Quark-Gluon Plasma (QGP)~\cite{BRAHMS:2004adc,STAR:2005gfr,PHENIX:2004vcz,PHOBOS:2004zne,ALICE:2022wpn}. The collective expansion of the QGP generates anisotropic particle emission patterns due to the geometry of the overlapping region of two colliding nuclei~\cite{Ollitrault:1992bk}. The anisotropy can be quantified by the anisotropic flow coefficients, $v_n$, defined as the coefficients of the Fourier expansion of the angular distribution of emitted particles~\cite{Voloshin:1994mz}:
\begin{align}
\frac{dN}{d\varphi} \propto 1 + 2\sum_{n=1}^{\infty}v_n\cos[n(\varphi-\Psi_n)]
\label{eq:fourier}
\end{align}
Anisotropic flow measurements at RHIC and the LHC have shown that the QGP behaves as a near-perfect liquid~\cite{STAR:2000ekf,PHENIX:2003qra,ALICE:2010suc,ALICE:2011ab,ALICE:2014wao,ALICE:2016ccg,ATLAS:2012at,ATLAS:2011ah,ATLAS:2013xzf,CMS:2012xss,CMS:2012zex}, with a shear viscosity over entropy density ratio close to the lowest limit predicted by AdS/CFT theory~\cite{Kovtun:2004de}. The anisotropic flow studies are typically performed via multi-particle azimuthal correlations~\cite{Borghini:2000sa}, providing insight into various aspects of QGP dynamics. These studies include anisotropic flow coefficients~\cite{STAR:2000ekf,PHENIX:2003qra,ALICE:2010suc,ALICE:2011ab,ALICE:2016ccg,ATLAS:2012at,CMS:2012xss}, flow coefficient fluctuations~\cite{ATLAS:2013xzf,CMS:2017glf,ALICE:2018rtz}, flow coefficient correlations~\cite{ATLAS:2015qwl,ALICE:2016kpq,ALICE:2021adw,STAR:2018fpo}, flow symmetry plane correlations~\cite{ATLAS:2014ndd,ALICE:2017fcd,ALICE:2019xkq,ALICE:2020sup,CMS:2019nct} and flow vector fluctuations/decorrelations~\cite{CMS:2013bza,CMS:2017mzx,ALICE:2017lyf,ALICE:2022dtx,ALICE:2024fcv,ATLAS:2023rbh}. As such, anisotropic flow studies are central to understanding the collective properties of the QGP~\cite{Voloshin:2008dg, Heinz:2013th, Shuryak:2014zxa, Song:2017wtw,Busza:2018rrf}. All such multi-particle correlations can be efficiently and precisely measured using the {\it Generic Framework}~\cite{Bilandzic:2013kga}, or its advanced version {\it Generic Algorithm}~\cite{Moravcova:2020wnf}. A critical aspect of these algorithms is their ability to properly incorporate detector efficiencies, ensuring accurate results for high-order correlations under realistic detector conditions. The $v_n$ coefficients ($n<=3$) are tightly correlated with the initial eccentricity, $\varepsilon_n$, of the initial state and is to first-order a linear hydrodynamic response to the initial state geometry~\cite{Niemi:2015qia}. This makes the anisotropic flow a valuable probe of the nuclear structure of the colliding nuclei~\cite{Jia:2022ozr}, which is imprinted in the initial geometry due to the low crossing time of the nuclei compared to collective degrees of freedom such as rotational and vibrational DOFs~\cite{Jia:2025wey}. For instance, the relative increase of the measured $v_2$ in central \XeXe collisions compared to central \PbPb collisions can only be explained by a sizable quadrupole deformation, $\beta_2 = 0.207$, of the $^{129}$Xe nuclei~\cite{ALICE:2024nqd}. 

Besides the study of anisotropic flow phenomenon using multi-particle azimuthal correlations, the mean transverse momentum, $\mpt$, and its fluctuations can also be studied using multi-particle transverse momentum correlations. They provide another approach to probe the initial conditions of heavy-ion collisions. More specifically, the $\mpt$ fluctuations, arising from the fluctuations in the initial size of the fireball~\cite{Broniowski:2009fm,Bozek:2012fw}, offer insight into how energy is deposited from the colliding nuclei into the system created in the early stages of the ultra-relativistic nuclear collisions~\cite{Giacalone:2020lbm} as they are strongly correlated to the initial energy density~\cite{Giacalone:2020dln}. The $\mpt$ fluctuations can be estimated by the fluctuations in the initial transverse size, $d_\perp$, of the system via $\dpt/\mpt = \delta d_\perp/d_\perp$~\cite{Niemi:2015qia,Schenke:2020uqq}. Measurements of the lower orders (mean and variance) of the event-by-event $\mpt$ fluctuations at RHIC~\cite{STAR:2005vxr} and LHC~\cite{ALICE:2014gvd} and comparison to theoretical models confirm that initial state density fluctuations are necessary to describe the observed $\mpt$ fluctuations. Measurements of higher-order moments of the $\mpt$ distribution, such as the skewness and kurtosis~\cite{ALICE:2023tej,ATLAS:2024jvf}, offer additional insight into the early stages of the ultra-relativistic nuclear collisions by constraining the initial-state fluctuations and even disentangling geometrical and intrinsic sources of fluctuations~\cite{Samanta:2023amp,Samanta:2023kfk}. The $\mpt$ fluctuations can be studied efficiently to arbitrary order using the generalised multi-particle $\pt$ correlations introduced in~\cite{Nielsen:2023znu}.

The correlation between $v_n^{2}$ and $\dpt$ probes the correlation between the initial size and shape, providing a more comprehensive probe of the initial state. This correlation, in the form of a modified Pearson Correlation Coefficient, $\rho(v_{2}^{2},\dpt)$~\cite{Bozek:2016yoj}, has been measured at RHIC~\cite{STAR:2024wgy} and the LHC~\cite{ALICE:2021gxt,ATLAS:2022dov}. The $\rho(v_{2}^{2},\dpt)$ has the distinct advantage that it is mostly insensitive to final state effects but instead directly reflects the correlations of the initial shape and size~\cite{ALICE:2021gxt}. The $\rho(v_{2}^{2},\dpt)$ calculations show a strong sensitivity to the nucleon width used in initial state models~\cite{Giacalone:2021clp} and was used to resolve a conflict in the description of the initial state between competing heavy-ion models~\cite{Nijs:2022rme}. 

Given the strong connection between $v_n$-$\dpt$ correlations and the initial-state geometry, these measurements serve as a key tool for studying nuclear structure at TeV energies. Due to the short crossing time of the colliding nuclei, heavy-ion collisions effectively capture an image of the overlap region between them. In central collisions, where the principal nuclear axes are aligned, this image reflects the ground-state nuclear structure. The anisotropic flow coefficients from two-particle correlations suffice to measure the quadrupole deformation, $\beta_2$; however, a nucleus such as $^{129}$Xe is expected to exhibit triaxiality, $\gamma$, where all three principal axes differ in length. Fully resolving such three-dimensional nuclear structures requires at least a three-particle cumulant. Comparison of measurements of the lowest order $v_n$-$\dpt$ three-particle correlation, $\rho(v_{2}^{2},\dpt)$ to theoretical calculations using the framework of energy-density functional methods~\cite{Bender:2003jk}, has suggested that the $^{129}$Xe nuclei are triaxially deformed with a fixed $\gamma = 27^\circ$~\cite{Bally:2021qys}. So far, nuclear structure studies in high-energy heavy-ion collisions have assumed rigid structures with fixed parameters of $\beta_2$ and $\gamma$ for $^{129}$Xe. However, $^{129}$Xe exists within a region of suspected phase transition on the nuclide chart. Within the Interacting Boson Model~\cite{Iachello_Arima_1987} and under the E(5) symmetry group~\cite{Iachello:2001ph,Iachello:2000ye}, the Xenon nuclei undergo a shape phase transition around $^{128\text{-}130}$Xe associated with $\beta$-soft and $\gamma$-soft deformation~\cite{Li:2010qu,Rodriguez-Guzman:2007fla,Fossion:2006xg,Nomura:2017ilh,Robledo:2008zz}. Higher-order moments of the joint $v_2$ and $\dpt$ distribution not only provide further insight into the initial-state properties of heavy-ion collisions but can also be used to explore the nuclear shape phase transition at the TeV energy scales~\cite{Zhao:2024lpc}. 

Traditionally, the lower-order correlation between azimuthal angle and transverse momentum has been measured using the sub-event method to suppress the non-collective contamination. Meanwhile, higher-order cumulants, which involve correlations among multiple particles, are expected to be less biased regarding such non-collective effects. Measuring these cumulants within the entire available phase space significantly increases the statistical precision by utilising the increased number of available particle tuples. This paper presents a general algorithm to obtain any correlation between arbitrary moments of a set of observables. Recursive formulae to generate the cumulants in specific cases are also presented. The algorithm is showcased through heavy-ion models such as HIJING and AMPT.

\section{Unified Algorithms for the Multi-Particle Correlations}\label{sec:algorithm} 
\subsection{One Algorithm to Rule Them All}\label{subsec:UA}

Two components are involved in computing multi-particle correlations. The first component is the computation of moments of the distribution of various observables. In general, this is $\la \mathcal{O}_1 \cdot \mathcal{O}_2 \cdots \mathcal{O}_N\ra$, where $\mathcal{O}_i$ is a specific observable.
Care must be taken when computing each observable, as each must be computed from a different particle. Doing so ensures that only dynamical fluctuations are considered and self-correlations are removed. This requirement, when applied naively, drastically increases the necessary computing time. Techniques are then applied to optimize this, but they become progressively more complicated as higher-order moments are investigated. The second component is to compute the actual correlation from the moments, $C(\mathcal{O}_1,\mathcal{O}_2,\cdots,\mathcal{O}_N)$. In general, this is done using the cumulant formulation\cite{Kubo}. Cumulants reveal whether a correlation is genuine in that it cannot be reduced to any lower-order set of correlations. Again, the formulae used to compute higher-order correlations become increasingly more complicated to calculate.

To compute multi-particle correlations, $C(\mathcal{O}_1,\mathcal{O}_2,\cdots,\mathcal{O}_N)$, we must first calculate the necessary moments, $M(\mathcal{O}_1,\mathcal{O}_2,\cdots,\mathcal{O}_N)$, defined in Eq.~\ref{eq:mom}.

\begin{equation}
 M(\mathcal{O}_1,\cdots,\mathcal{O}_N) = \frac{\displaystyle\sum_{i_1\neq\cdots\neq i_N}\omega_{1,i_1}\cdots\omega_{N,i_N}\mathcal{O}_{1,i_1}\cdots\mathcal{O}_{N,i_N}}{\displaystyle\sum_{i_1\neq\cdots\neq i_N}\omega_{1,i_1}\cdots\omega_{N,i_N}},
\label{eq:mom}
\end{equation}
The requirement that $i_1\neq\cdots\neq i_N$ creates considerable complication (necessary though to remove self-correlations). One must actually compute $2^N-1$ different sums of the various observables to remove these self-correlations without manually performing the calculation with nested loops. One must first fill an array (which we will call \verb|S| here) with all the various combinations of observables summed over all particles. This is, for example, $\sum_{i=1}^{N_{particles}}w_{j,i}w_{k,i}w_{l,i}\mathcal{O}_{j,i}\mathcal{O}_{k,i}\mathcal{O}_{l,i}$, if observables $j$, $k$, and $l$ are considered. Another array, \verb|SW|, contains only the weights for the denominator. To accomplish this, we use bits to represent whether a variable is or is not included in a sum. So \verb|S[0]=S[1-1]| is $\sum_{i=1}^{N_{particles}}w_{1,i}\mathcal{O}_{1,i}$, because 1 only has the lowest bit on. If we want $\sum_{i=1}^{N_{particles}}w_{1,i}w_{3,i}\mathcal{O}_{1,i}\mathcal{O}_{3,i}$, this is held in S[5-1] as 5 has the binary representation of 101. To fill \verb|S| and \verb|SW|, we use an array N elements long, called \verb|O| here, and fill it with each element being $\text{'weight'}\times\text{'observable'}$ for \verb|S| or just $\text{'weight'}$ for \verb|SW|. Then, the following C++ code can be executed for each considered particle in an event to fill \verb|S| and \verb|SW| from those arrays.
\begin{verbatim}
void fill_array(complex* S,
                complex* O,
                unsigned int n) {

  for (unsigned int i=(1<<n)-1; i>0; --i) {
    complex val = 1.; //assign 1 initially
    for (unsigned int iobs=0; iobs<n; ++iobs) {
      if ((i>>iobs)&1) val *= O[iobs];
    }
    S[i-1] += val;
  }
}
\end{verbatim}
Note that the \verb|complex| type has to be chosen from whatever is available in someone's code base and could change how initialization works.

Now that \verb|S| and \verb|SW| are filled, two new arrays, also with $2^{N}-1$ elements each, called \verb|M| and \verb|MW| are made to hold the moments with all self-correlations removed. The relevant recursion relation used to accomplish this is:
\begin{align*}
    \sum_{i_1\neq\cdots\neq i_N} & A_{1,i_1}\cdots A_{N,i_N} = \\ & \left(\sum_{i_N} A_{N,i_N}\right)\cdot\sum_{i_1\neq\cdots\neq i_{N-1}} A_{1,i_1}\cdots A_{N-1,i_{N-1}}\\ & -\sum_{i_1\neq\cdots\neq i_{N-1}} \left(A_{1,i_1}\cdot A_{N,i_1}\right)\cdot A_{2,i_2}\cdots A_{N,i_N}\\
    & \hspace{12mm}\vdots\\ & -\sum_{i_1\neq\cdots\neq i_{N-1}} A_{1,i_1}\cdots A_{N-2,i_{N-2}}\cdot \left(A_{N-1,i_{N-1}}\cdot A_{N,i_{N-1}}\right)\\
\end{align*}
This recursion relation must be applied many times to produce the algorithm. 
The following C++ code exploits this relationship in a slightly optimized way to accomplish this. One must run \verb|com_corrs(n,S,M)| and \verb|com_corrs(n,SW,MW)| after filling \verb|S| and \verb|SW| for all particles in the event, where \verb|n| is the number of observables.
\begin{verbatim}
unsigned int com_fac(unsigned int mask) {
  //Compute (n-1)! where n
  // is the number of set bits in mask
  unsigned int fac = 1, bit_count = 0;
  while (mask &= (mask-1)) fac *= (++bit_count);
  return fac;
}
\end{verbatim}
\vspace{2cm}
\begin{verbatim}
complex com_corr(unsigned int mask,
                 complex* S) {
  //Find min val with largest bit set
  unsigned int mask_hold = mask, maskmin = 1;
  while (mask_hold >>= 1) maskmin <<= 1;

  complex c = -1*com_fac(mask)*S[mask-1];
  for (unsigned int i=(mask-1)&mask;
       i>=maskmin;
       i=(i-1)&mask)
    c -= com_fac(i)*S[i-1]*com_corr((~i)&mask, S);
  return c;
}

void com_corrs(unsigned int nobs,
               complex* S,
               complex* M) {
  for (unsigned int i=(1<<nobs)-1; i>0; --i)
    M[i-1] = (i&1?-1:1)*com_corr(i, S);
}
\end{verbatim}

Now that \verb|M| and \verb|MW| are filled, the analyzer can decide how to average the moments over many events. This can be with unit weighting, multiplicity weighting, or whatever the analyzer chooses. The result averaged separately for each element in the array must end up in some final array (of length $2^N-1$), which we will call \verb|EM| here. Then, one can use the following code to calculate the cumulant from this by calling \verb|com_cumulant((1<<n)-1,EM)| where the first parameter produces a bit mask with all \verb|n| bits on. Note that \verb|(1<<n)-1| produces the value $2^N-1$.

\begin{verbatim}
complex com_cumulant(unsigned int mask,
                     complex* EM,
                     unsigned int depth=0) {
  //Find min val with largest bit set
  unsigned int mask_hold = mask, maskmin = 1;
  while (mask_hold >>= 1) maskmin <<= 1;

  complex c = 0;
  if ((mask-1)&mask)
    for (unsigned int i=(mask-1)&mask;
         i>=maskmin;
         i=(i-1)&mask)
      c += EM[i-1]*com_cumulant((~i)&mask,
                                EM,
                                depth+1);

  return EM[mask-1]-(depth+1)*c;
}
\end{verbatim}
This code is essentially a direct implementation of the following equation from \cite{Kubo}, also written in \cite{Bilandzic:2013kga}:
\begin{equation*}
    Cum(\{n\}) = \sum_{l=1}^n (l-1)!(-1)^l \sum_{\sum_{i=1}^l \{m_i\} = \{n\}} \prod_{i=1}^l Mom(\{m_i\})
\end{equation*}
where $\sum_{i-1}^l \{m_i\} = \{n\}$ represents all ways to divide $\{n\}$ in $l$ subsets and $Mom(\{m_i\})$ is the moment mentioned in Eq.~\ref{eq:mom} with the number of elements being $m_i$.

One can additionally define a new array \verb|C| of length $2^N-1$ and use the following code to compute that $N$ observable cumulant and all lower order cumulants.

\begin{verbatim}
void com_cumulants(unsigned int nobs,
                   complex* EM,
                   complex* C) {
  for (unsigned int i=(1<<nobs)-1;
       i>0;
       --i) 
    C[i-1] = com_cumulant(i, EM);
}
\end{verbatim}


\subsection{New observables of multi-particle cumulants of azimuthal angle and transverse momentum}\label{subsec:new_obs}


Among many different multi-particle correlations between azimuthal angle and transverse momentum, specific multi-particle cumulants of azimuthal angle and transverse moment (no flow symmetry plane is involved) that we will investigate here are:
\begin{align}
    &C(e^{in(\varphi_1-\varphi_2)},\dpt)   = C(v_n^2,\dpt) \\
    &= \la v_n^{2}\dpt\ra\nonumber\\
    &C(e^{in\varphi},e^{-in\varphi},\dpt_i,\dpt_j) = C(v_n^2,\dpt^2) \\
    &= \la v_n^{2}\dpt^{2}\ra - \la v_n^{2}\ra\la\dpt^{2}\ra\nonumber\\
    &C(e^{in(\varphi_1-\varphi_2)},\dpt_i,\dpt_j,\dpt_k) = C(v_n^2,\dpt^3) \\
    &= \la v_n^{2}\dpt^{3}\ra - 3\la v_n^{2}\dpt\ra\la\dpt^{2}\ra- \la v_n^{2}\ra \la\dpt^{3}\ra\nonumber\\
    &C(e^{in(\varphi_1+\varphi_2-\varphi_3-\varphi_4)},\dpt) = C(v_n^4,\dpt) \\ &= \la v_n^{4}\dpt\ra - 4\la v_n^{2}\dpt\ra\la v_n^{2}\ra\nonumber\\
    &C(e^{in(\varphi_1+\varphi_2-\varphi_3-\varphi_4)},\dpt_i,\dpt_j) = C(v_n^4,\dpt^2) \\
    &=\la v_n^{4}\dpt^{2}\ra - 4\la v_n^{2}\dpt^{2}\ra\la v_n^{2}\ra\nonumber +4\la v_n^{2}\ra^{2}\la \dpt^{2}\ra \\ &- 4\la v_n^{2}\dpt\ra^{2}-\la v_n^{4}\ra\la \dpt^{2}\ra\nonumber
\end{align}
All of these can be computed using the generic algorithm presented here. One should note that every power of $v_n$ or $\delta \pT$ represents a separate variable that must be computed from separate particles in the same event. Additionally, the notation of $v_n^p$, where $p$ is some even power, just represents the expected flow coefficient being measured. The actual observable for $v_n^2$ comes from one observable of $e^{i n \phi}$ and another of $e^{-i n \phi}$.
The cumulants can be normalised by some appropriate choice of denominator~\cite{Bozek:2021zim}. In this paper, the normalisation of $C(v_n^m,\dpt^k)$ is done by the $m$th-order anisotropic flow cumulant $c_n\{m\}$ and the $\pt$ variance, $\la\dpt^2\ra^{k/2}$
\begin{align}
    NC(v_n^2,\dpt) &= \frac{C(v_n^2,\dpt)}{c_n\{2\}\sqrt{\la\dpt^2\ra}}\label{eq:nc22pt}\\
    NC(v_n^2,\dpt^2) &= \frac{C(v_n^2,\dpt^2)}{c_n\{2\}\la\dpt^2\ra}\label{eq:nc22pt2}\\
    NC(v_n^2,\dpt^3) &= \frac{C(v_n^2,\dpt^3)}{c_n\{2\}\left(\la\dpt^2\ra\right)^{3/2}}\label{eq:nc22pt3}\\
    NC(v_n^4,\dpt) &= \frac{C(v_n^2,\dpt)}{c_n\{4\}\sqrt{\la\dpt^2\ra}}\label{eq:nc24pt}\\
    NC(v_n^4,\dpt^2) &= \frac{C(v_n^4,\dpt^2)}{c_n\{4\}\la\dpt^2\ra}\label{eq:nc24pt2}
\end{align}
The above choice will help to minimize the contamination of non-collective effects from azimuthal angle correlations.

\section{The models and setup}\label{sec:model}

The A Multi-Phase Transport model, AMPT~\cite{Lin:2004en}, is a widely used tool for studying high-energy nuclear collisions; it consists of four key components. The HIJING model~\cite{Wang:2000bf} simulates the spatial and momentum distributions of minijet partons and soft string excitations in the initial conditions. Then, Zhang’s Parton Cascade (ZPC) model~\cite{Zhang:1997ej} simulates the parton cascade, describing parton scatterings based on a screened two-body cross-section. The hadronization occurs through a quark-coalescence particle production mechanism~\cite{Chen:2005mr}, where nearby partons recombine into hadrons. In the end, the hadronic rescatterings are modelled using A Relativistic Transport (ART) model~\cite{Li:1995pra}, with the interaction strengths controlled via the time steps. The string melting version of AMPT is used in this paper, where the partonic degree of freedom is enabled. The AMPT-string melting model successfully reproduces particle production and anisotropic flow observed in heavy-ion collisions, which provides valuable inputs for the understanding of the transport properties of QGP~\cite{Alver:2010dn,Xu:2011fe,Bilandzic:2013kga,Zhou:2015eya,Nielsen:2022jms}. In addition, it captures the response of final-state collective expansion to the initial state properties (i.e., shape, size and their event-by-event fluctuations), making it a valuable tool for investigating the structure of the nuclei colliding at RHIC and the LHC~\cite{Giacalone:2021udy,Zhang:2021kxj,Jia:2022qgl,Nielsen:2023znu,Lu:2023fqd,Zhao:2024feh,Zhang:2024vkh}. In this paper, the multi-particle correlations (cumulants) between flow coefficients and transverse momentum, $C(v_n^m,\dpt^k)$, are investigated using AMPT simulations for Pb–Pb and Xe–Xe collisions at the LHC. Notably, calculations of higher-order correlations between azimuthal angle and transverse momentum are presented for the first time. These results can serve as baseline predictions, as they do not account for specific nuclear shapes of $^{208}$Pb or $^{129}$Xe.

The Heavy Ion Jet INteraction Generator model, HIJING, was developed to study hadron production in high-energy nucleon-nucleon, nucleon-nucleus, and nucleus-nucleus collisions~\cite{Wang:2000bf}. With additional modifications to the string configuration, this model can also describe bulk hadron spectra and high $p_{\rm T}$ hadron suppression in ultra-relativistic nuclear collisions~\cite{Deng:2010mv}. At the same time, HIJING has also been used as an initial condition generator for hydrodynamic models~\cite{Zhu:2016puf, Zhao:2017yhj, Zhao:2020pty} or parton cascade models like AMPT~\cite{Lin:2004en}. However, a key limitation of HIJING is its lack of collective effects, meaning it fails to reproduce the anisotropic flow measurements in ultra-relativistic nuclear collisions at RHIC and the LHC. Because of this, HIJING simulations serve as a baseline study where collective effects are absent, helping to investigate remaining experimental biases originating from non-collective effects~\cite{Bilandzic:2013kga, Huo:2017nms, Moravcova:2020wnf, Bhatta:2021qfk, Nielsen:2022jms}. This paper employs HIJING simulations to investigate potential biases arising from non-collective effects in the newly proposed multi-particle cumulants of flow coefficients and transverse momentum.

\begin{figure*}[t!]
\includegraphics[width=\textwidth]{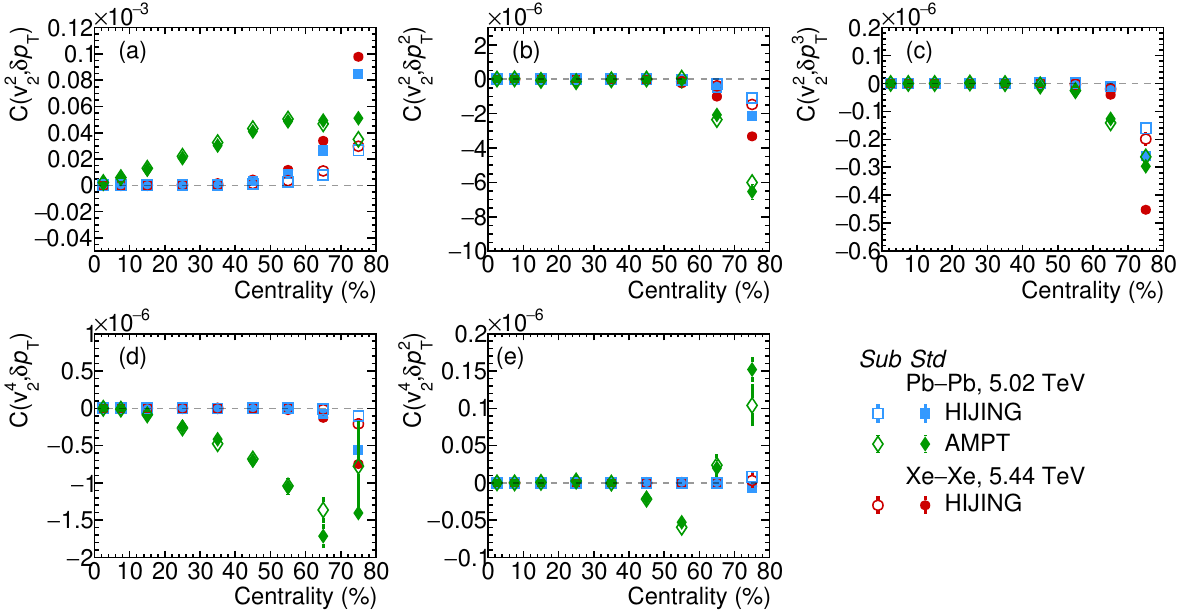}
\caption{\label{fig:multipanel} Centrality dependence of $C(v_{n}^{k}, \delta p_{\rm T}^{n})$ in Pb--Pb collisions at 5.02 TeV from AMPT (azure squares) and HIJING (green diamonds) models. Similar calculations in Xe--Xe collisions at 5.44 TeV from HIJING (red circles) are also presented. Calculations from the standard and sub-event methods are shown in solid and open markers, respectively.}
\end{figure*}

\section{Results and Discussion}\label{sec:results}

It has been established that the higher-order cumulants constructed in this work have the ability to provide unique insights into the structure of colliding nuclei. Their specific sensitivities to the quadrupole deformation $\beta_2$ and the triaxiality parameter $\gamma$ can be predicted by calculating an estimate from the liquid drop model, displayed in table \ref{tab:liquiddrop}. From this comparison, a statement can be made about the information that it is possible to extract from each cumulant.  

Generally, the degree of sensitivity that any given cumulant has to $\beta_2$ approximately scales with the number of total particles used in its calculation. However, the unique combinations of two- and four-particle azimuthal angle correlation components with different orders of $\dpt$ across cumulants induce different levels of dependence on $\gamma$. Notably, the four-particle cumulant $C(v_2^{2},\dpt^{2})$ is completely insensitive to $\gamma$ in the liquid drop formulation. The six-particle cumulant $C(v_2^{4},\dpt^{2})$ has the most significant sensitivity to the $\gamma$ fluctuations, making it particularly valuable in the context of the potential shape phase transition of the $^{129}$Xe nucleus. Due to its increased sensitivity, this cumulant should be sufficient to distinguish between a rigid structure with a fixed triaxiality and the $\gamma$ fluctuations that would indicate the phase transition.

\subsection{Baseline predictions for the collective behaviour}\label{subsec:ampt}

Most of the cumulants derived from the AMPT model have an overall negative trend, with the exception of the three-particle cumulant $C(v_2^{2},\dpt)$, which shows a positive correlation between $v_{2}^{2}$ and $\mpt$. For the non-normalized cumulant presented in figure~\ref{fig:multipanel}, $C(v_2^{2},\dpt)$ in panel (a) appears to have a strong dependence on centrality, increasing linearly before reaching a plateau around a centrality of 60\%. This behaviour is noted to not reflect in the normalized cumulants (see figure \ref{fig:allNorm} in appendix), which has a relatively weak dependence on centrality. There is, therefore, an indication that this strong centrality dependence is largely due to the contribution of the two-particle azimuthal angle correlation (or $v_{n}^{2}$) dominating the three-particle cumulant $C(v_2^{2},\dpt)$. 

This effect from the two-particle azimuthal angle correlations is not seen in the higher order cumulants $C(v_2^{2},\dpt^{2})$ (b) and $C(v_2^{2},\dpt^{3})$ (c), where an equal or greater number particles are included in the calculation of the cumulant's transverse momentum component. In these cumulants, very little centrality dependence is observed until the region of 60--80\%, where both undergo an exponential increase in magnitude with a negative trend. 

For the five-particle cumulant $C(v_2^{4},\dpt)$, shown in panel (d), the centrality dependence has almost mirrored the trend observed in $C(v_2^{2},\dpt)$ up to the 60\% region, but with a negative correlation. 
The magnitude is once again attributed to the domination of the four-particle azimuthal angle correlation in the five-particle cumulant. Unlike in the three-particle case, this five-particle cumulant does not go on to reach a plateau but instead reaches a peak between 60 and 70\%. 

Furthermore, the six-particle cumulant $C(v_2^{4},\dpt^{2})$, shown in panel (e), is of special interest due to its aforementioned sensitivity to the $\gamma$ fluctuations. In the AMPT model, it can be seen that $C(v_2^{4},\dpt^{2})$ has a negative trend in mid-central collisions but undergoes a change in sign in the centrality range of 60--70\% and increases exponentially for more peripheral collisions. There is clearly a strong dependence on centrality in this cumulant.

\subsection{Non-Collective Contamination}\label{subsec:bg}

As discussed in section~\ref{sec:model}, the HIJING model provides a baseline study that can be used to assess the contamination of the cumulant measurements from non-collective effects, as these are not accounted for within the model. 
This estimation applies exclusively to experimental measurements and should not be considered a dependable measure of non-collective contamination in the above-discussed AMPT calculations. The HIJING calculations show that all presented cumulants are consistent with zero in central and mid-central collisions.

In peripheral collisions, non-collective contributions start to present in the HIJING calculations. The magnitude of the contamination varies across the cumulants presented. In cumulants involving correlations of two azimuthal angle observables, a non-zero magnitude is always seen above a centrality of 60--70\%, similar to the standard two-particle azimuthal angle correlations.

Conversely, in the six-particle cumulant involving four-particle azimuthal angle correlations, the HIJING baseline is consistent with 0 across the full centrality range in the standard case. This demonstrates the additional suppression of non-collective effects associated with using a higher-order azimuthal angle correlation. It further suggests that the coming measurements of this cumulant will be largely unbiased due to non-collective effects across all centralities. This is particularly beneficial, as it provides motivation to use the larger sample accessible through the standard method in the search for potential $\gamma$ fluctuations using the six-particle cumulant.

Though it has been noted that the HIJING model is most relevant as an estimator of the non-collective contamination in experimental measurements, the presence of this exponential dependence on centrality in the HIJING baseline regions indicates a general contribution from non-collective effects in the peripheral regions which cannot be entirely disregarded in the AMPT model. These effects are possibly the source of the exponential increasing trend seen in the peripheral region of the cumulants obtained from the AMPT model.


The cumulants are also presented with and without applying the sub-event method. In the cumulants extracted from HIJING, the method is seen to have the expected effect, reducing the magnitude consistently across the centrality range. However, it must be noted that even when the sub-event method is included, the cumulants $C(v_2^{2},\dpt)$, $C(v_2^{2},\dpt^{2})$ and $C(v_2^{2},\dpt^{3})$ all maintain their divergence from zero in peripheral collisions. This suggests that the sub-event method does not succeed at removing all contributions from non-collective effects in this range of centrality. However, the sub-event method is almost entirely effective at suppressing the non-collective effects in the five-particle cumulant of $C(v_2^{4},\dpt)$.

\section{Summary}\label{sec:summary}

In this work, we have developed a {\it unified algorithm} to study multi-particle correlations between azimuthal angle and transverse momentum. This algorithm offers a precise and efficient method to compute correlations in any arbitrary order, which was not possible before. The technique provides a systematic approach to analyzing correlations in complex many-body systems, making it highly relevant for high-energy nuclear physics studies. We introduce several key observables involving multi-particle cumulants of flow coefficients and transverse momentum, denoted as \( C(v_n^m,\dpt^k) \). Based on the implementation of {\it unified algorithm}, these new observables have been validated using the AMPT and HIJING models. The AMPT results, exhibiting characteristic centrality dependence and sign changes, establish a baseline for studying the collective behaviour of these higher-order correlations. Meanwhile, the HIJING results and their deviations from zero serve as an essential reference for understanding non-collective effects in future experimental measurements.

Given the remarkable sensitivity of multi-particle correlations in azimuthal angle and transverse momentum to the initial conditions of heavy-ion collisions and the structure of the colliding nuclei, the proposed algorithm emerges as a powerful tool for investigating nuclear structure. Furthermore, it enables the exploration of nuclear shape phase transitions in high-energy nuclear collisions, opening new avenues for research at the energy frontier.

\section*{Acknowledgments}
This work is supported by the European Union (ERC, Initial Conditions), VILLUM FONDEN, with grant no. 00025462, and Danmarks Frie Forskningsfond (Independent Research Fund Denmark).


\appendix

\section{Additional plots}
\begin{figure*}
    \centering
    \includegraphics[width=\linewidth]{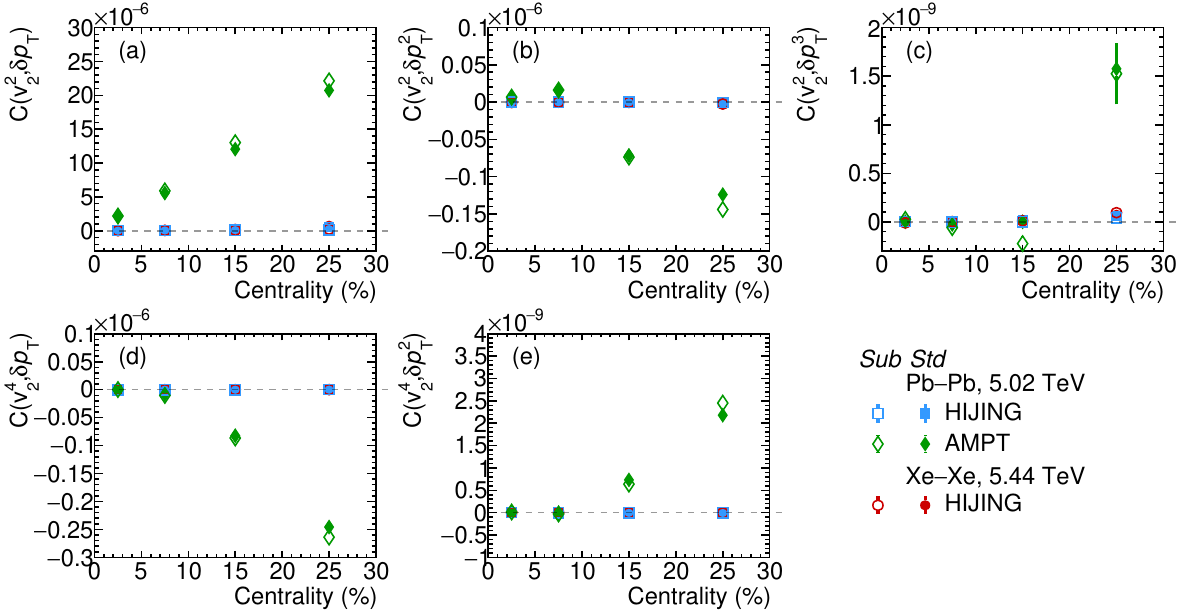}
    \caption{\label{fig:all030} Centrality dependence of $C(v_{n}^{k}, \delta p_{\rm T}^{n})$ in 0--30\% central Pb--Pb collisions at 5.02 TeV from AMPT (azure squares) and HIJING (green diamonds) models. Similar calculations in Xe--Xe collisions at 5.44 TeV from HIJING (red circles) are also presented. Calculations from the standard and sub-event methods are shown in solid and open markers, respectively.}
\end{figure*}

\begin{figure*}
    \centering    \includegraphics[width=\textwidth]{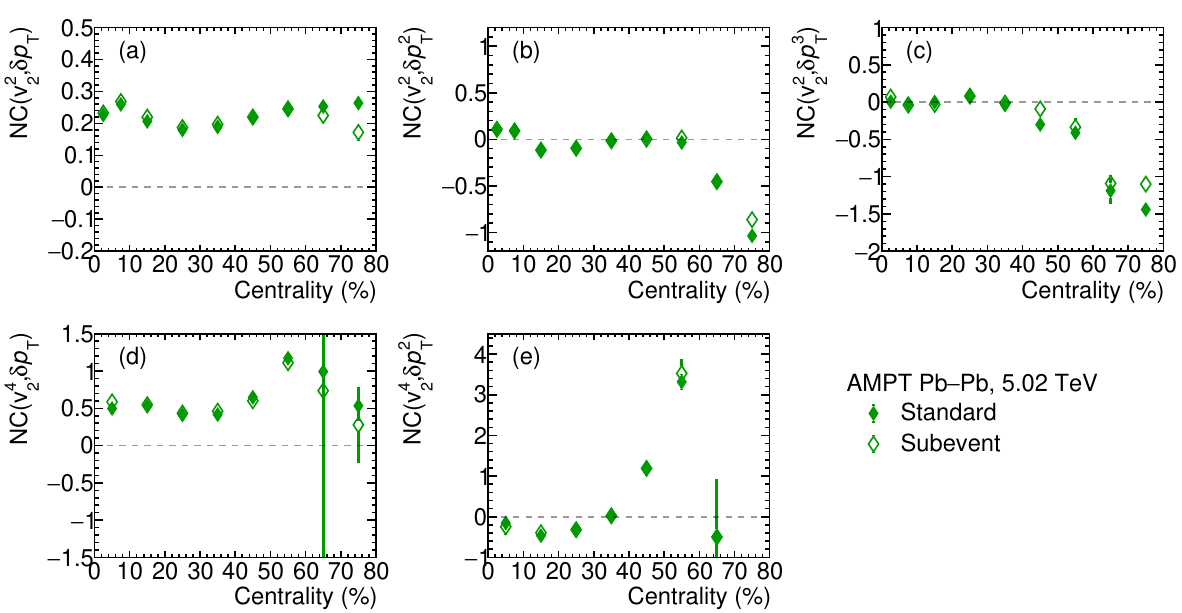}
    \caption{\label{fig:allNorm} Centrality dependence of $NC(v_{n}^{k}, \delta p_{\rm T}^{n})$ in Pb--Pb collisions at 5.02 TeV from AMPT (azure squares) and HIJING (green diamonds) models. Similar calculations in Xe--Xe collisions at 5.44 TeV from HIJING (red circles) are also presented. Calculations from the standard and sub-event methods are shown in solid and open markers, respectively.}
\end{figure*}

Figure \ref{fig:all030} shows the cumulants of flow coefficient and transverse momentum zoomed in 0-30\% centrality. Figure \ref{fig:allNorm} shows the normalized cumulants from Eq. \eqref{eq:nc22pt}-\eqref{eq:nc24pt2}.

\section{Liquid-drop model estimates}
The sensitivity of the multi-particle cumulants of flow coefficient and transverse momentum can be estimated by the liquid drop model \cite{Jia:2021qyu}. These estimates are shown for the observables presented in this paper in table \ref{tab:liquiddrop}.

\begin{table*}
\centering
\begin{tabularx}{\textwidth}{@{}c*1{>{\centering\arraybackslash}X}@{}*1{>{\centering\arraybackslash}X}@{}}
Final state	&Initial state	&Liquid-drop\\
cumulant    &cumulant   &model\\\hline
$C(v_{2}^{2},\dpt)$  &$\lla\varepsilon_{2}^{2}\frac{\delta d_\perp}{d_\perp}\rra$ &$-\frac{3\sqrt{5}\cos( 3\gamma) \la\beta_{2}^{3}\ra}{28\pi^{3/2}}$\\\hline
$C(v_{2}^{2},\dpt^{2})$ &$\lla\varepsilon_{2}^{2}\left(\frac{\delta d_\perp}{d_\perp}\right)^{2}\rra-\la \varepsilon_{2}^{2}\ra\lla\left(\frac{\delta d_\perp}{d_\perp}\right)^{2}\rra$   &$\frac{3\left(5\la\beta_{2}^{4}\ra-7\la\beta_{2}^{2}\ra^{2}\right)}{224\pi^{4}}$\\\hline
$C(v_{2}^{2},\dpt^{3})$ &$\lla\varepsilon_{2}^{2}\left(\frac{\delta d_\perp}{d_\perp}\right)^{3}\rra - \la \varepsilon_{2}^{2}\ra\lla\left(\frac{\delta d_\perp}{d_\perp}\right)^{3}\rra-3\lla\varepsilon_{2}^{2}\frac{\delta d_\perp}{d_\perp}\rra\lla\left(\frac{\delta d_\perp}{d_\perp}\right)^{2}\rra$   &$\frac{3\sqrt{5}\cos(3\gamma)\left(11\la\beta_{2}^{2}\ra\la\beta_{2}^{3}\ra-5\la\beta_2^{5}\ra\right)}{2464\pi^{5/2}}$\\\hline
$C(v_{2}^{4},\dpt)$ &$\lla\varepsilon_{2}^{4}\frac{\delta d_\perp}{d_\perp}\rra-4\la\varepsilon_{2}^{2}\ra\lla\varepsilon_{2}^{2}\frac{\delta d_\perp}{d_\perp}\rra$  &$\frac{9\sqrt{5}\cos(3\gamma)\left(11\la\beta_{2}^{2}\ra\la\beta_2\ra^{3}-5\beta_{2}^{5}\right)}{154\pi^{5/2}}$\\\hline
$C(v_{2}^{4},\dpt^{2})$ &$\lla\varepsilon_{2}^{4}\left(\frac{\delta d_\perp}{d_\perp}\right)^{2}\rra-\la\varepsilon_{2}^{4}\ra\lla\left(\frac{\delta d_\perp}{d_\perp}\right)^{2}\rra - 4\la\varepsilon_{2}^{2}\ra\lla\varepsilon_{2}^{2}\left(\frac{\delta d_\perp}{d_\perp}\right)^{2}\rra-4\lla\varepsilon_{2}^{2}\frac{\delta d_\perp}{d_\perp}\rra^{2}+4\la\varepsilon_{2}^{2}\ra^{2}\lla\left(\frac{\delta d_\perp}{d_\perp}\right)^{2}\rra$   &$\frac{3(42042\la\beta_{2}^{2}\ra^{3}-17160\cos(3\gamma)\la\beta_{2}^{3}\ra^{2}-45045\la\beta_{2}^{2}\ra\la\beta_{2}^{4}\ra+175(53+16\cos(6\gamma))\la\beta_2^{6}\ra}{32032\pi^{6}}$\\\hline
\end{tabularx}
\caption{Selected multi-particle cumulants of flow coefficients and transverse momentum in a liquid-drop model potential averaged over random orientations.}
\label{tab:liquiddrop}
\end{table*}



\bibliography{Reference}

\end{document}